\documentclass[journal]{IEEEtran}
\usepackage{amsmath,amsfonts}
\usepackage{algorithm}
\usepackage{algpseudocode}
\usepackage{array}
\usepackage[caption=false,font=normalsize,labelfont=sf,textfont=sf]{subfig}
\usepackage{textcomp}
\usepackage{stfloats}
\usepackage{url}
\usepackage{verbatim}
\usepackage{graphicx}
\usepackage{cite}
\usepackage{mathtools}
\usepackage{bbm}

\usepackage{xcolor}
\usepackage{textcomp}

\begin{document}

\title{Linear Periodically Time-Variant Digital PLL Phase Noise Modeling Using Conversion Matrices and Uncorrelated Upsampling}

\author{Hongyu Lu,~\IEEEmembership{Graduate Student Member,~IEEE}, Patrick P. Mercier,~\IEEEmembership{Senior Member,~IEEE}
\thanks{The authors are with the Department of Electrical and Computer Engineering, University of California at San Diego, La Jolla, CA 92093 USA. (e-mail: pmercier@ucsd.edu).}
}



\makeatletter
\def\ps@IEEEtitlepagestyle{%
  \def\@oddfoot{\mycopyrightnotice}%
  \def\@oddhead{\hbox{}\@IEEEheaderstyle\leftmark\hfil\thepage}\relax
  \def\@evenhead{\@IEEEheaderstyle\thepage\hfil\leftmark\hbox{}}\relax
  \def\@evenfoot{}%
}
\def\mycopyrightnotice{%
  \begin{minipage}{\textwidth}
  \centering \scriptsize
\textcopyright 2024 IEEE.  Personal use of this material is permitted.  Permission from IEEE must be obtained for all other uses, in any current or future media, including reprinting/republishing this material for advertising or promotional purposes, creating new collective works, for resale or redistribution to servers or lists, or reuse of any copyrighted component of this work in other works.
  \end{minipage}
}
\makeatother

\maketitle

\begin{abstract}
This paper introduces a conversion matrix method for linear periodically time-variant (LPTV) digital phase-locked loop (DPLL) phase noise modeling that offers precise and computationally efficient results to enable rapid design iteration and optimization. Unlike many previous studies, which either assume linear time-invariance (LTI) and therefore overlook phase noise aliasing effects, or solve LPTV systems with noise folding and multiple sampling rate conversions that heightens modeling and computational complexity, the proposed conversion matrix method allows the designer to represent LPTV systems using intuitive LTI-like transfer functions with excellent accuracy. Additionally, the uncorrelated upsampling method addresses the cross-correlated spectrum of cyclostationary noise sources by a simple matrix multiplication. This eliminates the need to consider the beat frequency of the upsampled noise source and the system with different sampling rates, thus improving computational efficiency. The proposed algorithm is applied to modeling a integer-$N$ DPLL with time-varying proportional loop gain, and the modeling accuracy is validated with Simulink transient simulations.

\end{abstract}


\section{Introduction}
Contemporary System-on-Chip (SoC) designs typically require the integration of multiple frequency synthesizers to deliver precise clocks for mixers and digital circuits. Such circuits can end up occupying large physical area (e.g., 25\% of the total chip area in \cite{LeeRF2019}), and can limit the performance of such systems. Digital phase-locked loops (DPLLs) are a compelling option for frequency synthesis, due to their technology-scalable area, high programmability, and excellent noise-power figure-of-merit (FoM) \cite{LeeRF2019,dartizioFractionalBangBangPLL2022,hwangLowJitterLowFractionalSpurRingDCOBased2022,buccoleri72fsTotalIntegratedJitterTwoCoreFractional2023,dartizio129to151GHz2022,kang6GHzDPLLUsing2023,seong320fsRMSJitter2019,dartizio767fslntegratedJitter712023,castoro25GHzDigitalPLL2023,add1,add2,add3}.

Designing high-performance DPLLs necessitates a comprehensive understanding of the output phase noise (PN). To determine the transfer function from each noise source to the output, many prior studies have treated PLLs as linear time-invariant (LTI) systems at the reference frequency, $f_\text{REF}$, neglecting noise aliasing caused by divider downsampling \cite{perrottModelingApproachSD2002,xuAnalysisDesignDigital2022,xuDesignMethodologyPhaseLocked2017,razaviDesignCMOSPhaseLocked2020,huChargeSharingLockingTechnique2022,dadaltLinearizedAnalysisDigital2008,huMultirateTimestampModeling2022a}. This LTI approximation is suitable for PLLs with relatively low bandwidth. However, for PLLs with very wide bandwidth to suppress the excessive PN from ring oscillators, e.g., 80 MHz in \cite{parkLowJitterRingDCOBasedFractionalN2022} and $\sim100$ MHz in \cite{jo135fsRmsJitter2023}, the aliased noise will contribute significantly to the in-band PN and thus cannot be ignored. Additionally, \cite{xuDesignMethodologyPhaseLocked2017} uses forward Euler mapping $z = 1+sT_\text{REF}$ for $z$-domain to $s$-domain conversion, making the model less accurate in predicting PN close to $f_\text{REF}/2$.

For precise modeling of PN in wide-bandwidth DPLLs and multiplying delay-locked loops (MDLLs), it's essential to consider the system as a discrete-time linear periodically time-variant (LPTV) system. The methodologies in \cite{syllaios2011,syllaiosLinearTimeVariantModeling2012b,santiccioliTimeVariantModelingAnalysis2019a,avalloneComprehensivePhaseNoise2021} provides great accuracy on their respective models under analysis. However, these math-intensive, topology-specific derivations hinder engineers from quickly adapting to increasingly complicated new topologies. The complexity increases further when a delta-sigma modulator (DSM) with a distinct clock frequency from the oscillator and reference is incorporated \cite{syllaiosLinearTimeVariantModeling2012b}, necessitating multiple sample-rate conversions within the loop. More importantly, these works can produce inaccurate results by ignoring the cross-correlation of the frequency-shifted spectrum, as the upsampled noise sequence is cyclostationary rather than wide-sense stationary.

The conversion matrix method, also known as the harmonic transfer matrix (HTM) method \cite{vanasscheSymbolicModelingPeriodically2002}, offers an alternative approach for solving transfer functions in LPTV systems with spectral correlation included. This approach, previously employed in mixer/$N$-path filter analysis \cite{hameedFrequencydomainAnalysisMixerfirst2015,hameedFrequencyDomainAnalysisPath2016,LuMixer2024}, has demonstrated simplified yet accurate results compared to traditional state-space equation solutions \cite{soerUnifiedFrequencyDomainAnalysis2010,ghaffariTunableHighQNPath2011,linAnalysisModelingGainBoosted2014,pavanAnalysisEffectSource2018}. Moreover, it has been leveraged in stability analysis of grid-connected PLLs in power converters \cite{chenImpedanceModellingStability2019,nianSequencesDomainImpedance2018,shahSmallSignalModelingDesign2020}. The primary advantage offered by the conversion matrix method is its capability to represent all LPTV components with matrices, where the elements denote the ratio of the input and output frequency-shifted spectrum. Consequently, formulating the problem is straightforward for designers, and computationally demanding matrix operations are offloaded to modern computers optimized for such calculations. The conversion matrix method provides a modeling framework that can be easily applied to different designs with arbitrary complexity.

\IEEEpubidadjcol

PLLs, which are inherently LPTV, appear to be an ideal candidate for the conversion matrix method. However, the incorporation of a DSM could significantly increase the computational complexity. To understand why, consider simulating the conversion gain of an $RC$-mixer using Spectre periodic steady-state (PSS) and periodic AC (PAC) analysis. When the clock frequency, $f_\text{s}$, is 1 GHz, the fundamental frequency in PSS equals to $f_\text{s} = 1$ GHz; the simulation can quickly converge and only a few sidebands, depending on the application, need to be saved. However, incorporating an additional mixer with a clock frequency $f_\text{s}'=f_\text{s} + \Delta f = $ 1.01 GHz changes the system's fundamental frequency to the beat frequency $\Delta f=$ 0.01 GHz. As a result, 100$\times$ more sidebands need to be saved, significantly increasing the simulation memory and time consumption. Similar challenges arise when using the conversion matrix method to model a PLL if the DSM frequency is not an integer multiple of the reference frequency. Here, the beat frequency becomes the fundamental frequency, potentially making the problem dimension of the conversion matrix method (similar to the number of sidebands in PSS) too large to be solved.

The contributions of this paper are as follows:
\begin{enumerate}
    \item A ground-up theory for discrete-time LPTV system noise analysis using the conversion matrix method, accommodating multiple different sampling rates for independent noise sources.
    \item Introducing uncorrelated upsampling that compensates for the cross-correlation of the frequency-shifted upsampled spectrum. It also allows ignoring any beat frequency between different-rate noise sources, thereby reducing computation time.
    \item Development of a modularized algorithm enabling easy adaptation to different PLL designs. The MATLAB code and simulation model can be downloaded from \cite{convmat_code}.
\end{enumerate}

The paper is organized as follows: Section II defines the conversion matrix method in discrete-time LPTV systems. Section III introduces uncorrelated upsampling and provides an example of modeling a PLL with time-varying proportional loop gain. Section IV compares the simulation results with the PN predicted by the model. Section V discusses the limitations of the proposed work. Finally, Section VI offers conclusions.

\section{Discrete-Time LPTV System Modeling}
\begin{figure}[!t]
\centering
\includegraphics[width=3.49in]{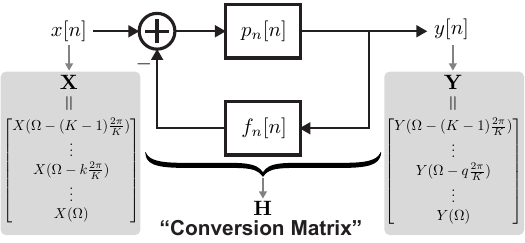}
\caption{Linear periodically time-variant (LPTV) system with input $x[n]$ and output $y[n]$.}
\label{fig:block_diagram_lptv_simple}
\end{figure}

In a discrete-time LPTV system depicted in Fig. \ref{fig:block_diagram_lptv_simple}, the relationship between input $x[n]$ and output $y[n]$ is characterized by the time-varying impulse responses of the feed-forward and feedback paths, represented as $p_n[n]$ and $f_n[n]$, respectively. The LPTV system's frequency conversion behavior, with a sampling rate of $f_\text{s} = 1/T_\text{s}$, can be expressed by a vector, $\mathbf{X}$, composed of the frequency-shifted input spectrum:

\begin{equation}
\mathbf{X}=
\begin{bmatrix} 
X(\Omega-(K-1)\frac{2\pi}{K}) \\ 
\vdots \\
X(\Omega-k\frac{2\pi}{K}) \\
\vdots \\
X(\Omega)
\end{bmatrix},
\label{eqn:def_X}
\end{equation}
in which $X(\Omega)$ is the discrete-time Fourier transform (DTFT) of $x[n]$ \cite[eq. (12.1)]{phillipsSignalsSystemsTransforms2014}, $\Omega = 2\pi fT_\text{s}$ is the discrete angular frequency, $K$ is defined as the \textit{problem dimension}, $k$ is the \textit{harmonic index} of the input, and $2\pi/K$ is the fundamental discrete angular frequency of the time-varying impulse response. 

Similarly, define the output vector, $\mathbf{Y}$, by replacing $X$ with $Y$ in \eqref{eqn:def_X}, 
\begin{equation}
\mathbf{Y}=
\begin{bmatrix} 
Y(\Omega-(K-1)\frac{2\pi}{K}) \\ 
\vdots \\
Y(\Omega-q\frac{2\pi}{K}) \\
\vdots \\
Y(\Omega)
\end{bmatrix},
\label{eqn:def_Y}
\end{equation}
in which $q$ is the harmonic index of the output. The output $\mathbf{Y}$ can be written as a matrix, $\mathbf{H}$, multiplied by the input $\mathbf{X}$:

\begin{equation}
\mathbf{Y}=\mathbf{H}\mathbf{X},
\label{eqn:tf_lptv}
\end{equation}
in which 
\begin{equation}
\mathbf{H}=
\begin{bmatrix} 
H_{K-1,K-1}&\hdots&H_{K-1,k}&\hdots&H_{K-1,0}\\
\vdots&\ddots&\vdots&&\vdots\\
H_{q,K-1}&\hdots&H_{q,k}&\hdots&H_{q,0}\\
\vdots&      &\vdots&\ddots&\vdots\\
H_{0,K-1}&\hdots&  H_{0,k}   &\hdots&H_{0,0}\\
\end{bmatrix}
\label{eqn:def_H}
\end{equation}
is defined as \textit{conversion matrix} that relates frequency-shifted input and output spectrum. The elements of $\mathbf{H}$, denoted as $H_{q,k}$, represent the gain from the $k^{\text{th}}$ input harmonic to the $q^{\text{th}}$ output harmonic. The discrete-time conversion matrix has a finite $K\times K$ dimension, as opposed to being a doubly infinite matrix in the continuous-time case \cite[eq. (12)]{vanasscheSymbolicModelingPeriodically2002}. This is because the DTFT of a discrete-time signal is $2\pi$-periodic, making the definitions in \eqref{eqn:def_X}-\eqref{eqn:def_H} contain all possible combinations of the input and output harmonics.

If the input $x[n]$ is a wide-sense stationary (WSS) stochastic process, the double-sideband power spectral density (PSD) of the output $y[n]$ can be written as \cite{levantinoFoldingPhaseNoise2010}:
\begin{equation}
S_{yy}(\Omega) = \sum_{k=0}^{K-1} |H_{0,k}(\Omega)|^{2}S_{xx}\left(\Omega-\frac{2\pi k}{K}\right).
\label{eqn:psd_out}
\end{equation}

The target now is to determine $\mathbf{H}$ for the system in Fig. \ref{fig:block_diagram_lptv_simple}, so \eqref{eqn:psd_out} can be used to calculate the output PSD. By inspection, 
\begin{equation}
\mathbf{Y} = \mathbf{P}(\mathbf{X} - \mathbf{F}\mathbf{Y}),
\label{eqn:tf_simple}
\end{equation}
in which $\mathbf{P}$ and $\mathbf{F}$ are conversion matrices associated with $p_n[n]$ and $f_n[n]$, respectively. Rearranging the terms in \eqref{eqn:tf_simple}, the output $\mathbf{Y}$ is 
\begin{equation}
\mathbf{Y} = \underbrace{[(\mathbf{I}+\mathbf{P}\mathbf{F})^{-1}\mathbf{P}]}_{\mbox{$\equiv\mathbf{H}$}}\mathbf{X},
\label{eqn:tf_simple_solution}
\end{equation}
in which $\mathbf{I}$ is a $K\times K$ identity matrix and $(\cdot)^{-1}$ denotes matrix inversion. Note that $\mathbf{H}$ in \eqref{eqn:tf_simple_solution} has a similar form as an LTI system transfer function with the loop gain defined as $\mathbf{L} = \mathbf{P}\mathbf{F}$. When implementing the algorithm, the direction of matrix multiplication needs to be consistent, as matrix multiplication is not commutative in most cases, i.e., $\mathbf{P}\mathbf{F}\neq\mathbf{F}\mathbf{P}$. With the definitions in \eqref{eqn:def_X}-\eqref{eqn:def_H}, signals always ``come in'' from the right side of the conversion matrix and ``leave from'' the left side.

Within an LPTV system, the input $x[n]$ can go through 2 types of building blocks: LTI subsystems, and multiplication with a periodic sequence. Subsequent sections will derive the conversion matrices for both blocks, ultimately leading to the calculation of $\mathbf{H}$.

\subsection{LTI Subsystem}
\begin{figure}[!t]
\centering
\includegraphics[width=3.49in]{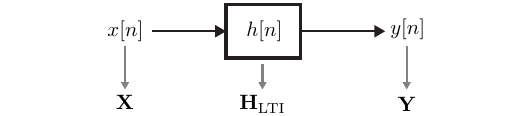}
\caption{Linear time-invariant (LTI) system with input $x[n]$ and output $y[n]$.}
\label{fig:block_diagram_lti}
\end{figure}

The discrete-time LTI system depicted in Fig. \ref{fig:block_diagram_lti} is characterized by a time-invariant impulse response, $h[n]$, and its corresponding DTFT, $H(\Omega)$. The relation between the input and output is given by:
\begin{equation}
Y(\Omega) = H(z)\bigg|_{z=e^{j\Omega}}X(\Omega),
\label{eqn:tf_lti}
\end{equation}
in which $H(z)$ is the $z$-transform of $h[n]$. By extending the scalar operation in \eqref{eqn:tf_lti} to the matrix operation described in \eqref{eqn:tf_lptv}, the output $\mathbf{Y}$ can be represented as:
\begin{equation}
\mathbf{Y}=
\underbrace{\begin{bsmallmatrix} 
H(\Omega-\frac{2\pi(K-1)}{K})&&&&\\
&\ddots&&&\\
&&H(\Omega-\frac{2\pi k}{K})&&\\
&&&\ddots&\\
&& &&H(\Omega)\\
\end{bsmallmatrix}}_{\mbox{$\equiv\mathbf{H}_\text{LTI}$}}\mathbf{X}.
\label{eqn:htf_lti}
\end{equation}
Here, $\mathbf{H}_\text{LTI}$ is the conversion matrix of the LTI system associated with the impulse response $h[n]$. The absence of frequency conversion in an LTI system renders $\mathbf{H}_\text{LTI}$ a $K\times K$ diagonal matrix.

\begin{figure}[!t]
\centering
\includegraphics[width=3.49in]{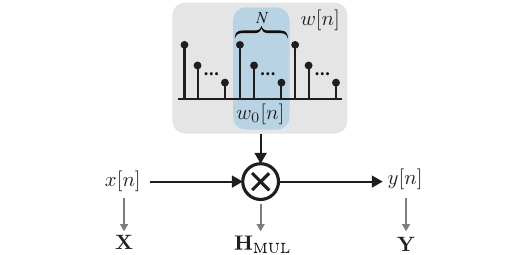}
\caption{Multiply the input $x[n]$ by a periodic sequence $w[n]$ and generate the output $y[n]$.}
\label{fig:block_diagram_mul}
\end{figure}

\subsection{Multiplication with Periodic Sequence}
Multiplying the input, $x[n]$, with a periodic sequence, $w[n]$, shown in Fig. \ref{fig:block_diagram_mul} is another fundamental block in LPTV systems. The sequence $w[n]$ has a period of $N$, i.e., $w[n] = w[n+N]$. Define $w_0[n]$ as one period of $w[n]$ \cite[eq. (12.17)]{phillipsSignalsSystemsTransforms2014}
\begin{equation}
w_{0}[n] = \begin{cases}
w[n],& 0\leq n\leq N-1 \\ 
{0},&{\text{else}} 
\end{cases}.
\end{equation}
The DTFT of $w[n]$ can be written as \cite[eq. (12.23)]{phillipsSignalsSystemsTransforms2014}
\begin{equation}
W(\Omega) = \frac{2\pi}{N}\sum_{k=-\infty}^{\infty} W_{0}\left(\frac{2\pi k}{N}\right) \delta\left(\Omega-\frac{2\pi k}{N}\right),
\end{equation}
in which $W_{0}(\Omega)$ is the DTFT of $w_0[n]$ \cite[eq. (12.1)]{phillipsSignalsSystemsTransforms2014}
\begin{equation}
W_{0}(\Omega) = \sum_{n=0}^{N-1} w_{0}[n] e^{-jn\Omega}.
\end{equation}

Multiplication in the time domain translates to circular convolution in frequency domain \cite[eq. (12.14)]{phillipsSignalsSystemsTransforms2014}. As a result, the output spectrum $Y(\Omega)$ can be written as
\begin{equation}
\begin{aligned}
Y(\Omega) &= \textbf{DTFT}\{x[n]w[n]\}\\ 
&= \frac{1}{N}\sum_{k=-\infty}^{\infty} W_{0}\left(\frac{2\pi k}{N}\right) X\left(\Omega-\frac{2\pi k}{N}\right).
\end{aligned}
\label{eqn:htf_mul_basis}
\end{equation}
Using \eqref{eqn:htf_mul_basis}, the output $\mathbf{Y}$ becomes
\begin{equation}
\begin{aligned}
\mathbf{Y}=
\underbrace{\begin{bmatrix} 
H_{0}&\hdots&H_{1-j}&\hdots&H_{1-N}\\
\vdots&\ddots&\vdots&&\vdots\\
H_{i-1}&\hdots&H_{0}&\hdots&H_{i-N}\\
\vdots&      &\vdots&\ddots&\vdots\\
H_{N-1}&\hdots&  H_{N-j}   &\hdots&H_{0}\\
\end{bmatrix}}_{\mbox{$\equiv\mathbf{H_\text{MUL}}$}}\mathbf{X},
\end{aligned}
\end{equation}
where $\mathbf{H}_\text{MUL}$ is the conversion matrix corresponding to the multiplication of the periodic sequence with the input. The element of $\mathbf{H}_\text{MUL}$ in the $i^{\text{th}}$ row and $j^{\text{th}}$ column is given by
\begin{equation}
H_{\text{MUL}(i,j)}=\frac{1}{N}W_{0}\left(\frac{2\pi (i-j)}{N}\right).
\label{eqn:mul_element}
\end{equation}
Like its continuous-time counterpart, $\mathbf{H}_\text{MUL}$ is a $K\times K$ frequency-independent Toeplitz matrix \cite[eq. (27)]{vanasscheSymbolicModelingPeriodically2002}.

\section{PLL Modeling Example}
With the derived building blocks, this section models a digital PLL with fast phase-error correction (FPEC) \cite{seong320fsRMSJitter2019} as an example. In contrast to a conventional PLL that maintains a constant proportional loop gain $K_\text{P0}$, FPEC boosts $K_P$ at the beginning of each reference cycle to rapidly eliminate the residue jitter from the previous reference cycle \cite[Fig. 2]{seong320fsRMSJitter2019}, resulting in reduced average jitter. This is a time-varying concept, and cannot be modeled with conventional LTI techniques. 

\begin{figure}[!t]
\centering
\includegraphics[width=3.49in]{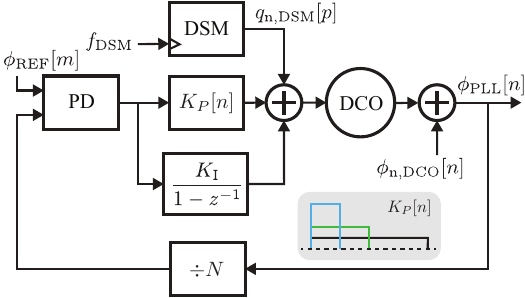}
\caption{High-level block diagram of the PLL with time-varying proportional loop gain $K_P[n]$.}
\label{fig:block_diagram_PLL_behav}
\end{figure}

The high-level block diagram of the Type-II DPLL with FPEC is shown in Fig. \ref{fig:block_diagram_PLL_behav}. The $N$-periodic time varying loop gain $K_P[n]$ can be written as:
\begin{equation}
K_{P}[n] = \begin{cases}
\frac{NK_\text{P0}}{P},& kN\leq n\leq kN+P-1 \\ 
{0},&{\text{else}} 
\end{cases},
\label{eqn:Kp[n]}
\end{equation}
in which $k\in\{0,1,2,\hdots\}$, $P\in[1,N]$ is the length of the FPEC window, and $K_\text{P0}$ is the proportional loop gain when FPEC is deactivated, i.e., $P = N$. When $P$ decreases, the height of $K_P[n]$ increases while keeping the total area within one reference period the same. Subsequent subsections will derive the DCO-rate transfer functions for noise sources with different sampling rates.

The second-order DSM used in this work adopts the structure depicted in \cite[Fig. 8]{dsmgalton} and connects its input to the integral path similar to \cite[Fig. 10]{seong320fsRMSJitter2019} to enhance the DCO tuning resolution. The DSM has 2 delays for the input control code and 2 zeros at the origin for the quantization noise. Unless otherwise mentioned, 2 simplifications relative to \cite{seong320fsRMSJitter2019} are employed:
\begin{enumerate}
    \item The proportional path and integral path directly control the capacitor banks with the same resolution digitally.
    \item All bits in the integral path are connected to the DSM input. Fractional resampling and 2 delays in the signal transfer function (STF) are ignored. 
\end{enumerate}
The second simplification may lead to inaccurate results because the fractional resampling of the control code \cite{syllaiosLinearTimeVariantModeling2012b} and the delays are ignored. However, the introduced error is negligible because $K_I \ll K_P$ in practical cases. To avoid distracting the reader from the main framework, the second simplification is made and will be revisited in Section \ref{fracsamp}.

\subsection{Noise Sampling Rate Conversion} \label{sec:rate_conv}
All noise sources must be converted to the DCO rate before being fed into the DCO-rate PLL model. As depicted in Fig. \ref{fig:block_diagram_PLL_behav}, the noise sources operate at 3 different rates:
\begin{enumerate}
  \item DCO noise, $\phi_\text{n,DCO}[n]$, running at $f_\text{DCO}$, which is the highest rate in the model.
  \item DSM quantization noise, $q_\text{n,DSM}[p]$, running at $f_\text{DSM} = f_\text{DCO}/M$. In most cases $1\leq M<N$ for pushing the shaped quantization noise to higher frequency.
  \item Reference noise, $\phi_\text{n,REF}[m]$, running at $f_\text{REF} = f_\text{DCO}/N$, which ideally should be the fundamental frequency of the model, making $K=N$ in \eqref{eqn:def_X}. The phase detector (PD) quantization noise can be lumped into the reference noise by dividing PD gain $K_\text{PD}$.
\end{enumerate}
Among these sources, DCO noise can be directly fed into the DCO-rate PLL model, while the sampling rate of the reference and DSM noise necessitates upconversion to the DCO rate.

\begin{figure}[!t]
\centering
\includegraphics[width=3.49in]{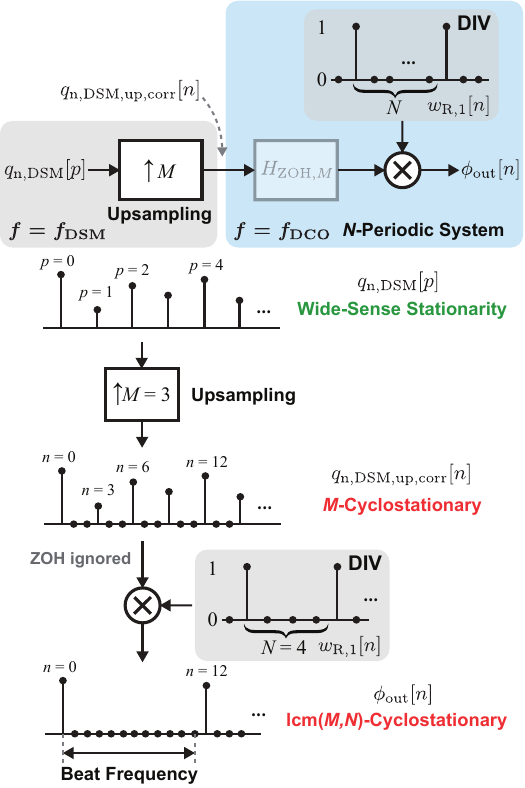}
\caption{Cyclostationary noise introduced by upsampling a WSS noise sequence by $M$ times. Beat frequency can be generated in the $N$-periodic system if $N$ is not an integer multiple of $M$. The zero-order hold is ignored only for simpler illustration of the beat frequency; it's included in the final model in Fig. \ref{fig:block_diagram_math}.}
\label{fig:cyclo_intro}
\end{figure}

Proper handling of sample-rate conversion is crucial for calculating transfer functions in LPTV systems. An example of upsampling the DSM noise, $q_\text{n,DSM}[p]$, is shown in Fig. \ref{fig:cyclo_intro} with $M=3$ and $N=4$. After upsampling, $q_\text{n,DSM,up,corr}[n]$ is no longer WSS but $M$-cyclostationary, i.e., its mean $\mu[n]=\mu[n+lM]$ and autocorrelation sequence $R_{qq}[n,k] = R_{qq}[n+lM,k]$ are periodic over $M$, where $l$ is an integer. When $q_\text{n,DSM,up,corr}[n]$ passes through an $N$-periodic system, the PSD of the output, $\phi_\text{out}[n]$, cannot be calculated by \eqref{eqn:psd_out} anymore because it only applies to WSS input that doesn't have cross-correlated frequency-shifted spectrum \cite[eq. (23)]{vanasscheSymbolicModelingPeriodically2002}.

However, previous works on LPTV PLL analysis use \eqref{eqn:psd_out} on a cyclostationary input while still demonstrating excellent matching between calculations and simulations. Based on observation, \eqref{eqn:psd_out} generates no or little error for an $N$-periodic system with $M$-cyclostationary input in one of the following cases:
\begin{enumerate}
    \item $N$ and $M$ are coprime, i.e., $\text{gcd}(N,M)=1$. In the example showed earlier, $N = 4$ and $M = 3$, \eqref{eqn:psd_out} can be used because $\text{gcd}(3,4)=1$. In prior works that model DSM noise, \cite{syllaios2011} uses $N=25$ and $M=4$, and \cite{syllaiosLinearTimeVariantModeling2012b} uses $N = 43$ which is a prime number. This is similar to the continuous-time case described in \cite[III-A]{cyclo2000}.
    \item $M = N$, and the $M$-cyclostationary input comes from upsampling a WSS white noise sequence that has a flat time-averaged PSD.
\end{enumerate}

\begin{figure}[!t]
\centering
\includegraphics[width=3.49in]{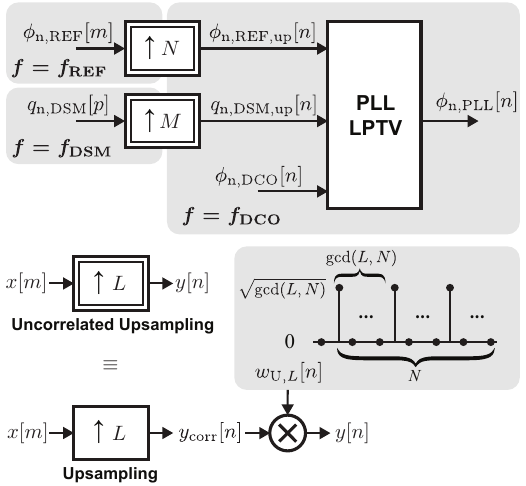}
\caption{Multi-rate LPTV PLL model for time-varying loop gain analysis. All noise sources are converted to DCO rate with uncorrelated upsampling for further processing.}
\label{fig:block_diagram_math_pre}
\end{figure}

The special case 1) has an interesting interpretation: when $\text{gcd}(N,M)=1$, the $N$-periodic system treats its $M$-cyclostationary input as WSS, so \eqref{eqn:psd_out} applies. In other words, the system can’t distinguish whether its input is WSS or not. To generalize special case 1) to all $M$ and $N$ combinations, the sample-rate upconversion in this work is achieved through \textit{uncorrelated upsampling}, as shown in Fig. \ref{fig:block_diagram_math_pre}. Uncorrelated upsampling by a factor of $L$ involves upsampling by inserting $L-1$ zeros between adjacent samples of the input $x[m]$, followed by multiplying a decorrelation sequence, $w_{\text{U},L}[n]$, that removes correlated frequency spectrum with respect to the fundamental discrete angular frequency, $2\pi/N$. It ``tricks'' the $N$-periodic system into interpreting its input as WSS regardless of $L$.

The PSD of the upsampled sequence $y_\text{corr}[n]$, $S_{yy,\text{corr}}(\Omega)$, can be written as
\begin{equation}
S_{yy,\text{corr}}(\Omega) = \frac{1}{L}S_{xx}\biggl(\text{mod}(L\Omega,2\pi)\biggr), \Omega \in [0,2\pi),
\label{eqn:upsample}
\end{equation}
in which $\text{mod}(a,b)$ finds the remainder after division of $a$ by $b$. One period of $w_{\text{U},L}[n]$ is expressed as
\begin{equation}
w_{\text{U0},L}[n] = \begin{cases}
\sqrt{\mathrm{gcd}(L,N)},& \text{if mod}(n,\mathrm{gcd}(L,N))=0 \\ 
{0},&{\text{else}} 
\end{cases},
\end{equation}
in which $0\leq n\leq N-1$, and $\mathrm{gcd}(L,N)$ computes the greatest common divisor of $L$ and $N$. The decorrelation sequence $w_{\text{U},L}[n]$ is the $N$-periodic extension of $w_{\text{U0},L}[n]$:
\begin{equation}
w_{\text{U},L}[n] = w_{\text{U0},L}[\text{mod}(n,N)],
\label{eqn:decorrelationwindow}
\end{equation}
and its associated conversion matrix $\mathbf{W}_{\text{U},L}$ can be found with \eqref{eqn:mul_element}. An intuitive explanation of $w_{\text{U},L}[n]$ is shown in the Appendix. Note that multiplying the upsampled noise source with a decorrelation sequence $w_{\text{U},L}[n]$ does not occur in an actual PLL; it only has mathematical significance, compensating for the underestimated output PSD due to the cross-correlated frequency-shifted spectrum of the cyclostationary input.

Another practical problem that uncorrelated upsampling solves is that it allows using $f_\text{REF}$ as the fundamental frequency of the system. This is because the cyclostationary input appears WSS to the system after decorrelation. Otherwise, any beat frequency lower than $f_\text{REF}$ becomes the fundamental frequency, as shown in Fig. \ref{fig:cyclo_intro}, unnecessarily enlarging the problem dimension as $K$ increases in \eqref{eqn:def_X} to capture all possible frequency conversions. For example, if $N=51$ and $M=50$, the problem dimension $K$ becomes $\text{lcm}(M,N) = 2550$, where lcm represents the least common multiple. Considering the $O(n^3)$ complexity when inverting an $n\times n$ matrix, the proposed uncorrelated upsampling reduces the computation time by $(\text{lcm}(M,N)/N)^3 = 125000\times$.

\subsection{DCO-Rate Transfer Function Analysis}
\begin{figure*}[!t]
\centering
\includegraphics[width=7.14in]{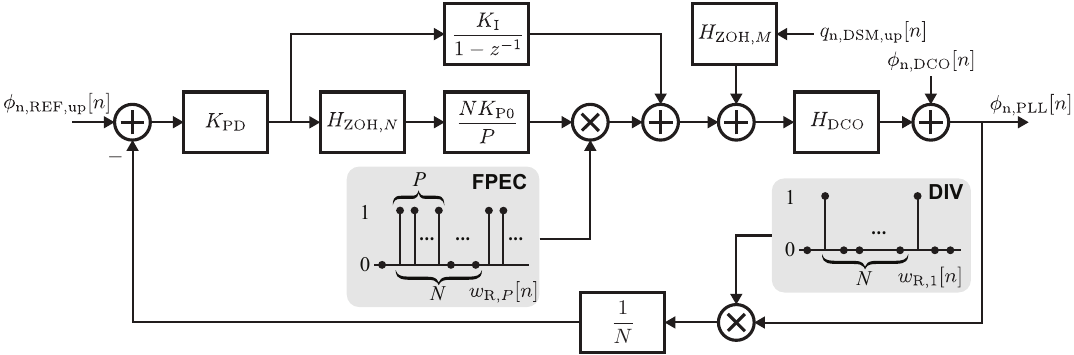}
\caption{DCO-rate LPTV model for the FPEC PLL. The zero-order hold (ZOH) block $H_{\text{ZOH},N}$ is used to keep the value of the feedback signal constant during one reference period $T_\text{REF}$, and $H_{\text{ZOH},M}$ holds the upsampled DSM noise constant during $T_\text{DSM}$.}
\label{fig:block_diagram_math}
\end{figure*}
After converting all noise sources to the DCO rate, noise transfer functions are derived utilizing Fig. \ref{fig:block_diagram_math}. The divider is modeled as multiplying the output phase by a $N$-periodic pulse, $w_{\text{R},1}[n]/N$, that samples the output phase every $N$ samples:
\begin{equation}
w_{\text{R},1}[n] = \begin{cases}
1,& n = kN \\ 
{0},&{\text{else}} 
\end{cases}
\label{eqn:wr1}
\end{equation}
for $k\in\{0,1,2,\hdots\}$. The time-varying loop gain implements \eqref{eqn:Kp[n]} by multiplying a constant $NK_\text{P0}/P$ with a $N$-periodic rectangular window $w_{\text{R},P}[n]$:
\begin{equation}
w_{\text{R},P}[n] = \begin{cases}
1,& kN\leq n\leq kN+P-1 \\ 
{0},&{\text{else.}} 
\end{cases}
\label{eqn:wrp}
\end{equation}
for $k\in\{0,1,2,\hdots\}$. Multiplying with \eqref{eqn:wr1} or \eqref{eqn:wrp} falls under the case depicted in Fig. \ref{fig:block_diagram_mul}, and the associated conversion matrices,$\mathbf{W}_{\text{R},1}$ and $\mathbf{W}_{\text{R},P}$, can be calculated using \eqref{eqn:mul_element}.

The conversion matrices for the following LTI transfer functions can be obtained using \eqref{eqn:htf_lti}:
\begin{enumerate}
    \item $\mathbf{H}_{\text{ZOH},N}$: conversion matrix of $N$-length zero-order hold $H_{\text{ZOH},N}(z)=\frac{1-z^{-N}}{1-z^{-1}}$, where $z = e^{j\Omega}$.
    \item $\mathbf{K}_\text{I}$: conversion matrix of the integral path transfer function $\frac{K_\text{I}}{1-z^{-1}}$.
    \item $\mathbf{H}_\text{DCO}$: conversion matrix of the DCO transfer function $H_\text{DCO}=\frac{K_\text{DCO}}{f_\text{DCO}}\frac{z^{-1}}{1-z^{-1}}$ \cite{syllaiosLinearTimeVariantModeling2012b}. $K_\text{DCO}$ is the DCO gain with unit Hz/LSB.
\end{enumerate}

The conversion matrices for $\phi_\text{n,ref,up}[n]$, $q_\text{n,DSM,up}[n]$, and $\phi_\text{n,DCO}[n]$ to the output $\phi_\text{n,PLL}[n]$ are
\begin{equation}
\begin{split}
\mathbf{H}_\text{REF,tot} &= (\mathbf{I}-\mathbf{L})^{-1}\mathbf{P}_\text{REF}\\
\mathbf{H}_\text{DSM,tot} &= (\mathbf{I}-\mathbf{L})^{-1}\mathbf{P}_\text{DSM}\\
\mathbf{H}_\text{DCO,tot} &= (\mathbf{I}-\mathbf{L})^{-1}\mathbf{P}_\text{DCO}
\end{split}
\label{eqn:tftot}
\end{equation}
respectively, in which $\mathbf{L}$ is the LTI-like loop gain
\begin{equation}
\mathbf{L}=-K_\text{PD}\mathbf{H}_\text{DCO}\underbrace{\left(\frac{NK_\text{P0}}{P}\overbrace{\mathbf{W}_{\text{R},P}}^{\text{FPEC}}\mathbf{H}_{\text{ZOH},N} + \mathbf{K}_\text{I}\right)}_{\text{Loop Filter}}\overbrace{\frac{\mathbf{W}_{\text{R},1}}{N}}^{\text{Divider}}
\label{eqn:loopgain}
\end{equation}
and
\begin{equation}
\begin{split}
\mathbf{P}_\text{REF} &= K_\text{PD}\mathbf{H}_\text{DCO}\left(\frac{NK_\text{P0}}{P}\mathbf{W}_{\text{R},P}\mathbf{H}_{\text{ZOH},N} + \mathbf{K}_\text{I}\right)\\
\mathbf{P}_\text{DSM} &= \mathbf{H}_\text{DCO}\mathbf{H}_{\text{ZOH},M}\\
\mathbf{P}_\text{DCO} &= \mathbf{I}
\end{split}
\label{eqn:pathgain}
\end{equation}
are path gains from each input noise source to the output. Together with \eqref{eqn:psd_out} and uncorrelated upsampling shown in Fig. \ref{fig:block_diagram_math_pre}, the output phase noise, $\mathcal{L}(\Omega)$, can be determined as \cite{galtonUnderstandingPhaseError2019}:
\begin{equation}
    \mathcal{L}(\Omega) = \frac{1}{2}S_{\phi,\text{PLL}}(\Omega),
\end{equation}
in which $S_{\phi,\text{PLL}}(\Omega)$ is the SSB PSD of $\phi_\text{n,PLL}[n]$.

The process of finding the PLL output phase noise is summarized in Algorithm \ref{alg1}.
\begin{algorithm}[H]
\caption{Finding Output PN Using Conversion Matrix.}\label{alg:alg1}
\textbf{Input:} $S_{xx}(\Omega)$, $L$ \\
\textbf{Output:} $\mathcal{L}(\Omega)$
\begin{algorithmic}
\State $S_{yy,\text{corr}}(\Omega) \gets S_{xx}(\Omega) \uparrow L$ \Comment{Upsampling by $L$, \eqref{eqn:upsample}}
\State Compute $\mathbf{W}_{\text{R},1}$, $\mathbf{W}_{\text{R},P}$, $\mathbf{W}_{\text{U},L}$ \Comment{Window CM, \eqref{eqn:mul_element}}
\For{$i = 1:\text{length}(\mathbf{\Omega}_\text{vec})$}
\State $\Omega_0 \gets \mathbf{\Omega}_\text{vec}(i)$
\State Evaluate $\mathbf{H}_\text{LTI}(\Omega_0)$ \Comment{\eqref{eqn:htf_lti}}
\State $\mathbf{L}(\Omega_0) \gets f(\mathbf{W}_R,\mathbf{H}_\text{LTI}(\Omega_0)) $ \Comment{Loop gain, \eqref{eqn:loopgain}}
\State $\mathbf{P}(\Omega_0) \gets g(\mathbf{W}_R,\mathbf{H}_\text{LTI}(\Omega_0)) $ \Comment{Path gain, \eqref{eqn:pathgain}}
\State $\mathbf{H}(\Omega_0) \gets (\mathbf{I}-\mathbf{L}(\Omega_0))^{-1}\mathbf{P}(\Omega_0)$ 
\State $\mathbf{H}_{\text{U},L}(\Omega_0) \gets \mathbf{H}(\Omega_0)\mathbf{W}_{\text{U},L}$ \Comment{Decorrelation, \eqref{eqn:decorrelationwindow}}
\State $\mathcal{L}(\Omega_0)\gets h(\mathbf{H}_{\text{U},L}(\Omega_0),S_{yy,\text{corr}}(\Omega_0))$ \Comment{\eqref{eqn:psd_out}}
\EndFor
\end{algorithmic}
\label{alg1}
\end{algorithm}

In Algorithm \ref{alg1}, the left arrow ``$\leftarrow$'' represents the assignment of the evaluated expression on the right side to the variable on the left, $\mathbf{\Omega}_\text{vec}$ contains frequency values of interest sampled within $[0,\pi)$, and the algorithm solves a $N\times N$ linear system of equations each time with all LTI transfer functions evaluated at $\Omega_0 = \mathbf{\Omega}_\text{vec}(i)$.

\section{Simulation Results}
The proposed algorithm, which calculates the output PN using conversion matrices, is validated with Simulink transient simulations. The Simulink model implements Fig. \ref{fig:block_diagram_math} with parameters shown in Table \ref{tab:sim_param}. The TDC resolution, $\Delta t_\text{res}$, is set to 15.16 ps, leading to $K_\text{PD}=1/(2\pi f_\text{REF}\Delta t_\text{res})$ = 300 $\text{rad}^{-1}$ referring to the reference frequency $f_\text{REF} = 35$ MHz. The reference is assumed to have much less phase noise than the TDC, and is thus neglected in the following simulations.

\begin{table}[!t] 
\caption{Simulink Simulation Parameters\label{tab:sim_param}}
\centering
\begin{tabular}{|c|c|}
\hline
\textbf{Parameter} & \textbf{Value} \\ \hline 
N                 & 18             \\ \hline
$K_\text{PD}$          & 300 $\text{rad}^{-1}$           \\ \hline
$K_\text{P0}$          & 0.4            \\ \hline
$K_\text{I}$             & $K_\text{P0}/32$    \\ \hline
$K_\text{DCO}$         & 4 MHz/LSB      \\ \hline
$f_\text{REF}$         & 35 MHz         \\ \hline
\end{tabular}
\end{table}

\subsection{DCO}
\begin{figure}[!t]
\centering
\includegraphics[width=3.49in]{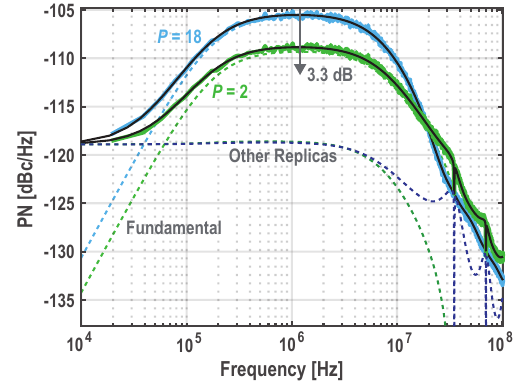}
\caption{Simulated output PN due to DCO noise with FPEC window length $P=2$ and without FPEC ($P=18$).}
\label{fig:sim_dco_pn}
\end{figure}

Fig. \ref{fig:sim_dco_pn} illustrates results when only the DCO noise source is active. The DCO noise is modeled as a white noise passing through a discrete-time integrator. The calculated output PN matches the simulated PN extremely well for both $P = N = 18$ and $P = 2$, including dips located at integer multiples of the reference frequency. The FPEC can suppress the in-band PN by 3.3 dB, while the PN above 20 MHz is increased.

\begin{figure}[!t]
\centering
\includegraphics[width=3.49in]{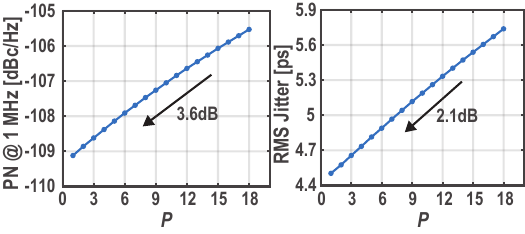}
\caption{PN and jitter improvement with different FPEC window length $P$. RMS jitter is obtained by integrating PN from 10 kHz to $f_{\text{DCO}}/2$.}
\label{fig:sim_dco_pn_inband_jitter}
\end{figure}

The effect of FPEC window length $P$ on in-band PN and total integrated jitter is shown in Fig. \ref{fig:sim_dco_pn_inband_jitter}. When $P$ reduces from 18 to 1, the in-band phase noise reduces 3.6 dB with a slope of 0.21 dB, and the RMS jitter reduces from 20.7 ps to 16.2 ps with a slope of 0.26 ps.

\subsection{DSM}
\begin{figure}[!t]
\centering
\includegraphics[width=3.49in]{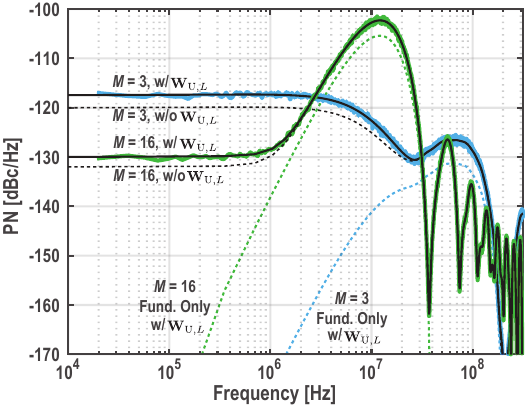}
\caption{Simulated output PN due to DSM quantization noise for DSM clock divide ratio $M=3$ and $M=16$. The FPEC window length $P$ is the same as reference clock divide ratio $N$ = 18.}
\label{fig:sim_dsm_uuwin}
\end{figure}

The PLL output PN due to DSM quantization noise is shown in Fig. \ref{fig:sim_dsm_uuwin}. The DSM quantization noise is modeled as a white noise with power of $1/12$ passing through $(1-z^{-1})^{2}$. For both $M=3$ and $M=16$ cases, the proposed model, shown in the solid black lines, precisely predict the simulated PN shown in the blue and green lines. The decorrelation sequence \eqref{eqn:decorrelationwindow} corrects the 3 dB lower in-band PN shown in the dashed black lines where the decorrelation sequence is not used. The high-frequency PN shows approximately 1 dB improvement in accuracy as well.

\begin{figure}[!t]
\centering
\includegraphics[width=3.49in]{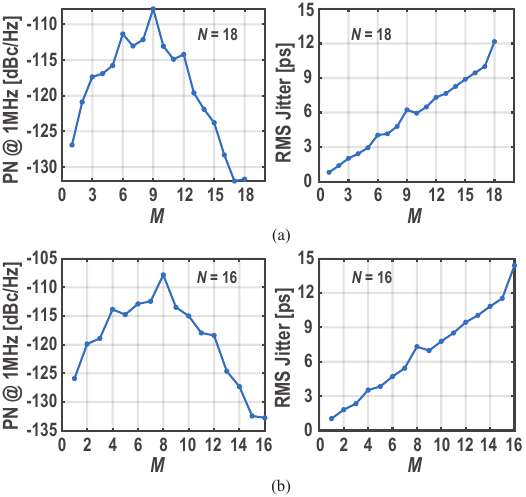}
\caption{Simulated output PN and output RMS jitter due to DSM quantization noise with different DSM clock divide ratio $M$ for (a) $N$ = 18 and (b) $N$ = 16.}
\label{fig:sim_dsm_pn_inband_jitter}
\end{figure}

Contrary to the continuous-time analysis predicting that the DSM noise has a high-pass shape (similar to the dashed green and blue lines) in the output PN, the actual low-frequency PN is dominated by aliased DSM noise. The output PN at 1 MHz and the integrated jitter is shown in Fig. \ref{fig:sim_dsm_pn_inband_jitter} with different DSM clock divide ratio, $M$, for $N$ = 18 and $N$ = 16. With smaller $M$, i.e., larger $f_\text{DSM}$, the quantization noise is pushed to higher frequency, resulting in a general trend of smaller RMS jitter. The 1 MHz spot PN, however, first increases with larger $f_\text{DSM}$, peaks at $M/2$, then decreases. When the in-band PN is critical in the application, $M$ needs to be carefully selected to avoid excessive noise aliasing. If the above results were to be generated without uncorrelated upsampling, the time consumption will be much larger under certain cases, e.g., 125$\times$ for $N=18$ and $M=5$.

Certain $M$ values lead to higher aliased noise, as seen in the peaks of Fig. \ref{fig:sim_dsm_pn_inband_jitter}. To find these numbers, we can factor $N$ into multiplication of prime numbers. For example, $N = 18 = 2\times3\times3$. If $M$ can be factored to prime numbers within the same set $\{2,3\}$, then this $M$ will generate higher PN and jitter due to aliasing. For example, $M = 2,3,4,6,9,10,12,14,15,16,18$. Among these numbers, $M = 3,6,9,12,15$ generates an even higher noise floor, as 3 occurs twice when factoring $N$. Similarly, for $N = 16 = 2\times2\times2$, any even $M$ generates more aliased PN. These $M$ values are preferred to be avoided when choosing the DSM operating frequency given a reference divide ratio $N$.

\begin{figure}[!t]
\centering
\includegraphics[width=3.49in]{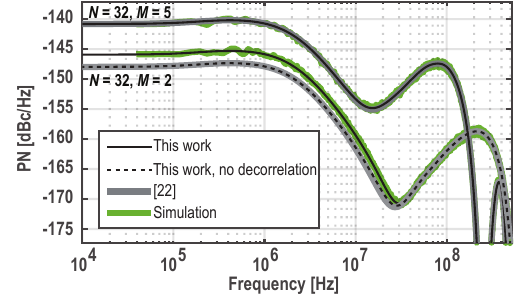}
\caption{Simulated output PN due to DSM quantization noise with $M = 2,5$ and $N=P=32$. The proposed model generates correct results regardless of $M$ and $N$, while \cite{syllaios2011} is only accurate when $\text{gcd}(M,N) = 1$.}
\label{fig:dsm_gcd}
\end{figure}

Fig. \ref{fig:dsm_gcd} compares the accuracy of the proposed work with \cite{syllaios2011}. When $N=32$ and $M = 5$, \cite{syllaios2011} generates the same accurate result as this work because $\text{gcd}(M,N) = 1$, which falls into the special case 1) mentioned in Section \ref{sec:rate_conv}. When $M = 2$, $\text{gcd}(M,N) = 2$, making \cite{syllaios2011} predict 2.1 dB lower close-in aliased PN than the simulation, which is the same as this work but without the decorrelation sequence $w_{\text{U},L}[n]$.

\subsection{TDC}
\begin{figure}[!t]
\centering
\includegraphics[width=3.49in]{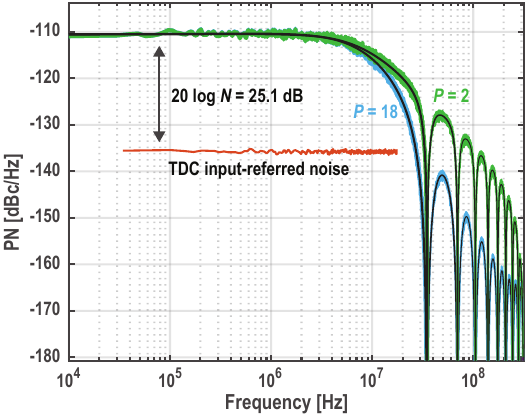}
\caption{Simulated output PN due to TDC quantization noise for FPEC window length $P=2$ and without FPEC ($P=N=18$).}
\label{fig:sim_tdc_pn}
\end{figure}

The simulated output PN due to TDC quantization noise is shown in Fig. \ref{fig:sim_tdc_pn} for FPEC window length $P=2$ and in the absence of FPEC ($P=18$). The TDC input-referred quantization noise floor is $\mathcal{L}_\text{TDC,in} = 1/(12)/K_\text{PD}^{2}/f_\text{REF} = -135.8$ dBc/Hz. The PLL output PN has a low-pass shape with a low-frequency plateau $20\log N = 25.1$ dB higher than the TDC input-referred noise floor. The colored simulation results matches well with the black lines predicted by the model. In the case of $P=2$, the output jitter due to TDC is 3.43 ps, 1 dB higher than 3.04 ps when $P=18$. 

\begin{figure}[!t]
\centering
\includegraphics[width=3.49in]{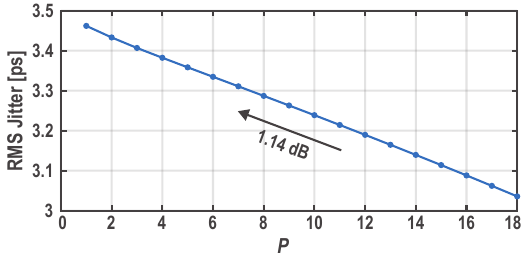}
\caption{Output RMS jitter due to TDC quantization noise for different FPEC window length $P$.}
\label{fig:sim_tdc_jitter}
\end{figure}

Fig. \ref{fig:sim_tdc_jitter} shows the output RMS jitter due to TDC quantization noise for different FPEC window length $P$. Using FPEC with shorter pulse width $P$ increases the reference noise contribution while suppressing more DCO noise, as shown in Fig. \ref{fig:sim_dco_pn}. This trade-off is leveraged by using an optimal-threshold TDC (OT-TDC) \cite{seong320fsRMSJitter2019} that generates less quantization noise by providing 2 more quantization thresholds than the bang-bang PD with optimal spacing.

\begin{figure}[!t]
\centering
\includegraphics[width=3.49in]{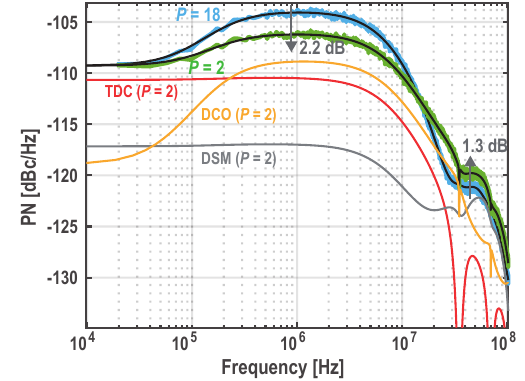}
\caption{Total output PN with TDC quantization noise, DSM quantization noise, and DCO phase noise for $N=18$, $M=4$, and $P = 18,2$.}
\label{fig:sim_tot_pn}
\end{figure}

\begin{figure}[!t]
\centering
\includegraphics[width=3.49in]{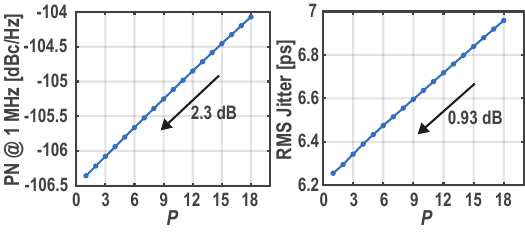}
\caption{Output PN and RMS jitter with TDC quantization noise, DSM quantization noise, and DCO phase noise for different $P$ under $N=18$ and $M=4$.}
\label{fig:sim_tot_pn_inband_jitter}
\end{figure}

\subsection{Total Output PN}

Fig. \ref{fig:sim_tot_pn} shows the output PN with all noise sources activated: TDC quantization noise, DSM quantization noise, and DCO phase noise. When $P$ reduces from 18 to 2, the in-band PN shows a 2.2 dB improvement, while the out-of-band noise plateau due to DSM quantization noise rises by 1.3 dB. Fig. \ref{fig:sim_tot_pn_inband_jitter} shows the in-band PN and RMS jitter improvement in finer resolution of $P$. The integrated RMS jitter improves only 0.93 dB when $P$ reduces from 18 to 2, a result of the trade-off between lower DCO noise and higher reference/TDC noise. When using a mid-rise TDC, the linearized PD gain becomes dependent on the RMS jitter shown at its input. Therefore, the curve of the output RMS jitter vs. $P$ becomes convex while remaining monotonic.

\begin{figure}[!t]
\centering
\includegraphics[width=3.49in]{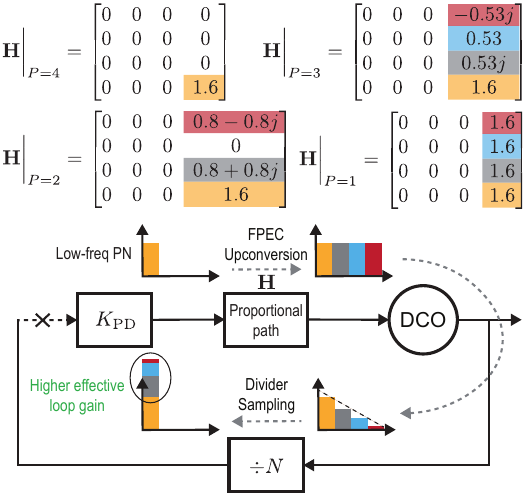}
\caption{Intuitive explanation of FPEC in the frequency domain using conversion matrix. The proportional path is composed of $\text{ZOH}_N$ and $K_P[n]$.}
\label{fig:fpec_explain}
\end{figure}

An intuitive explanation of FPEC in the frequency domain is given as follows: Fig. \ref{fig:fpec_explain} shows the conversion matrix of the loop filter without the integral path for $K_{\text{P0}}=0.4$, $N=4$, and $1\leq P \leq 4$ ($P=4$ disables FPEC). When $P=4$, the conversion matrix of the proportional path $\mathbf{H}$ only provides gain at the baseband (no frequency conversion). When $P=1$, FPEC creates identical copies of the baseband PN at higher frequencies. These upconverted PN, after being filtered by the DCO, alias back to the baseband due to divider sampling, which adds on top of the original PN. This effectively increases the loop gain, helping to suppress DCO noise, as shown in Fig. \ref{fig:sim_dco_pn}. In other words, FPEC suppresses in-band noise by ``seeking help'' from the out-of-band gain.

This DPLL can also be understood as a loopback transceiver \cite{loopback}, where the proportional path with FPEC functions as the mixer in the transmitter, the DCO serves as the power amplifier, the divider acts as a mixer-first receiver, and the PD works as an error amplifier in the baseband. The FPEC enables ``carrier aggregation'' \cite{LeeRF2019}, where multiple ``channels'' shown as colored rectangles in Fig. \ref{fig:fpec_explain} are used simultaneously to transmit the baseband data (PN), thereby increasing the received data strength.

\subsection{DSM Fractional Resampling and Delay} \label{fracsamp}

\begin{figure}[!t]
\centering
\includegraphics[width=3.49in]{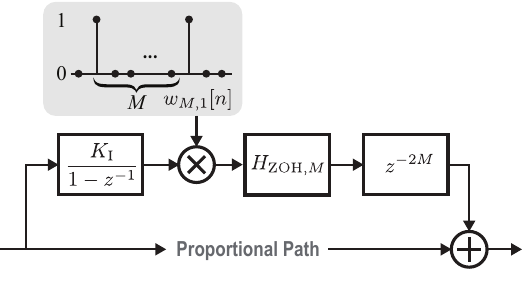}
\caption{Loop filter in Fig. \ref{fig:block_diagram_math} with modified integral path considering the DSM control code fractional resampling and delay.}
\label{fig:frac_resamp_model}
\end{figure}

This subsection revisits the effect of fractional resampling and delay in the integral path introduced by the DSM. The modified loop filter that takes them into account is shown in Fig. \ref{fig:frac_resamp_model}. Following the integrator, the DSM resampling sequence $w_{M,1}[n]$ with period $M$ is multiplied. An $M^{\text{th}}$-order ZOH is used to keep the value constant within one period, and finally, 2 delays at $f_\text{DSM}$ are converted to $z^{-2M}$ at $f_\text{DCO}$. The elements in the resampling conversion matrix $\mathbf{H}_\text{MUL}$ can be derived from \eqref{eqn:htf_mul_basis} and \eqref{eqn:mul_element} as follows:

\begin{equation}
H_{\text{MUL}(i,j)}=\frac{1}{\text{lcm}(N,M)}W_{0}\left(\frac{2\pi (i-j)}{N}\right),
\label{eqn:frac_samp_element}
\end{equation}
in which
\begin{equation}
W_{0}(\Omega) = \sum_{n=0}^{\text{lcm}(N,M)-1} w_{M,1}[n] e^{-jn\Omega}.
\label{eqn:frac_samp_window}
\end{equation}

\begin{figure}[!t]
\centering
\includegraphics[width=3.49in]{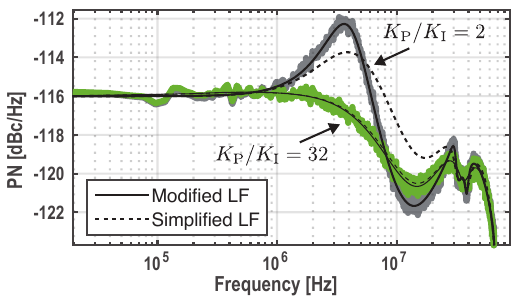}
\caption{Simulated output PN due to DSM quantization noise for $N=18$, $M=5$, and $P=4$. The simplified loop filter in Fig. \ref{fig:block_diagram_math} generates negligible error for $K_\text{P}/K_\text{I}=32$.}
\label{fig:frac_resamp_sim_kpki}
\end{figure}

Fig. \ref{fig:frac_resamp_sim_kpki} compares the results of the simplified loop filter in Fig. \ref{fig:block_diagram_math} and the modified one in Fig. \ref{fig:frac_resamp_model} for $K_\text{P}/K_\text{I}=32$ and $K_\text{P}/K_\text{I}=2$. In practical cases where $K_\text{I} \ll K_\text{P}$, it is safe, depending on the required accuracy, to ignore the dynamics in the integral path except for the integrator itself.

\begin{figure}[!t]
\centering
\includegraphics[width=3.49in]{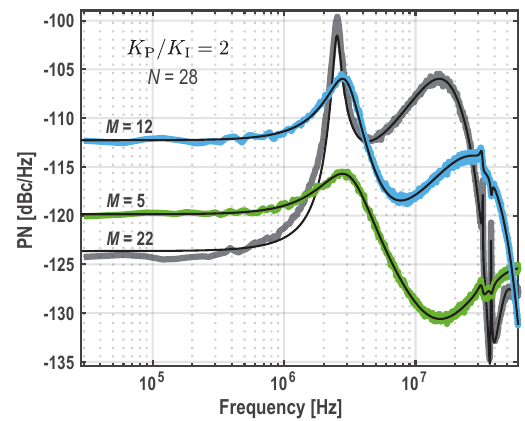}
\caption{Simulated output PN due to DSM quantization noise for $N=28$, $P=4$, and $M=5,12,22$ considering DSM fractional resampling.}
\label{fig:frac_resamp_sim_diffM}
\end{figure}

Note that \eqref{eqn:frac_samp_element} and \eqref{eqn:frac_samp_window} do not provide an exact solution for fractional resampling as they ignore some of the frequency shifts introduced by the beat frequency between $f_\text{REF}$ and $f_\text{DSM}$. Fig. \ref{fig:frac_resamp_sim_diffM} shows the simulated output PN due to DSM quantization noise with different $M$. $K_\text{P}/K_\text{I}$ is set to 2 to not attenuate the error from the integral path. Based on observations, using \eqref{eqn:frac_samp_element} gives negligible error for all $M \leq N/2$, and the error starts to increase when $M$ approaches $N$ while still being relatively small.

\subsection{Cross Verification}
\begin{figure}[!t]
\centering
\includegraphics[width=3.49in]{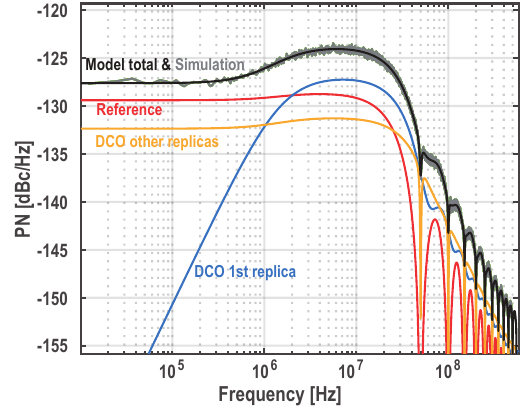}
\caption{Simulated output PN of an MDLL shown in \cite{santiccioliTimeVariantModelingAnalysis2019a} and calculated PN using the proposed model.}
\label{fig:cross_verify}
\end{figure}

Fig. \ref{fig:cross_verify} replicates the result of an MDLL modeled in \cite[Fig. 14]{santiccioliTimeVariantModelingAnalysis2019a} using the proposed method to demonstrate flexibility. Once the framework is built, the only work left to reproduce the result is rewriting the LTI-like transfer functions \eqref{eqn:tftot}-\eqref{eqn:pathgain} by inspection according to the block diagram shown in \cite[Fig. 11]{santiccioliTimeVariantModelingAnalysis2019a}.

\begin{figure}[!t]
\centering
\includegraphics[width=3.49in]{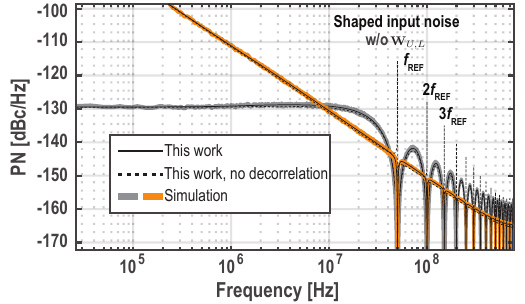}
\caption{Simulated output PN of an MDLL shown in \cite{santiccioliTimeVariantModelingAnalysis2019a} due to colored and white reference noise. The proposed work generates correct predictions under both cases, while removing the decorrelation sequence generates large spikes at integer-multiple of $f_\text{REF}$ when the reference noise is colored.}
\label{fig:m20dbref}
\end{figure}

Fig. \ref{fig:m20dbref} explores the validity of the proposed method with reference noise being white and colored ($-$20 dB/dec). While this work generates correct predictions in both cases, removing the decorrelation sequence generates large error spikes at integer multiples of $f_\text{REF}$ for colored input noise. This is a result of violating special case 2) and applying \eqref{eqn:psd_out} on an $N$-cyclostationary upsampled sequence, similar to Fig. \ref{fig:dsm_gcd}. This effect can be avoided by assuming white reference noise and ignoring the frequency shift \cite[eq. (30)]{santiccioliTimeVariantModelingAnalysis2019a}, \cite[eq. (41)]{avalloneComprehensivePhaseNoise2021}, then applying the result to colored reference noise, which ultimately gives negligible error.

\begin{figure}[!t]
\centering
\includegraphics[width=3.49in]{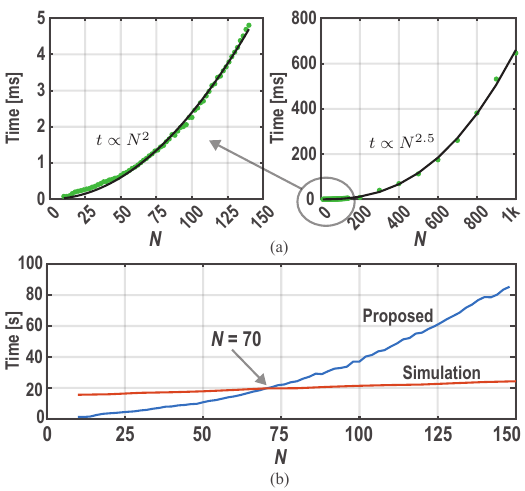}
\caption{(a) Computation time for one frequency-domain sample with different problem dimensions. The green dots are the actual computation time, and the fitted curves are shown in black. (b) Comparison of computation time for $2^{14}$ frequency-domain samples between the proposed work and the simulation. The number of periodogram averages is set to 500 for the simulation.}
\label{fig:sim_time}
\end{figure}

\subsection{Algorithm Complexity}
Fig. \ref{fig:sim_time}(a) shows the computation time required for a single frequency-domain sample using the proposed conversion matrix-based algorithm across varying problem dimensions $N$, specifically when $M = N-1$ and $P=2$. This algorithm was evaluated on a laptop equipped with an AMD Ryzen 9 5900HS CPU and 16.0 GB of RAM. For values where $N<150$, computation time exhibits a relationship proportional to $N^2$. However, for greater dimensions ($N\geq200$), the complexity appears to be proportional to $N^{2.5}$. The proposed uncorrelated upsampling approach can reduce computation time by a factor of at least $(N-1)^2$. When $N<65$, the computation time for one frequency-domain sample remains under 1 ms, and consumes less time than the simulation for $N<70$ as shown in Fig. \ref{fig:sim_time}(b). The efficiency of the algorithm
can be further leveraged if only a few, instead of $2^{14}$, spot
PN needs to be optimized, allowing the designer to quickly
explore the design space to fine-tune PN at selected frequencies and study noise folding with great accuracy.

\section{Limitations}
The proposed work has the following limitations:
\begin{enumerate}
    \item Linear circuits. In the case of a bang-bang PLL, this work can serve as a analysis tool to study noise folding, with the linear BBPD/OT-TDC gain extracted from either a behavioral transient simulation or a time-domain jitter analysis \cite{xuDesignMethodologyPhaseLocked2017,kang6GHzDPLLUsing2023,kennedytdc}.
    \item Integer-$N$ PLLs. Fractional-$N$ operation makes $f_\text{DCO}$ a non-integer multiple of $f_\text{REF}$, which cannot be easily captured by the proposed model. 
\end{enumerate}

\section{Conclusion}
This paper introduces the conversion matrix method into modeling a discrete-time LPTV system, such as DPLLs. The conversion matrices for two fundamental building blocks, namely the LTI subsystem and multiplication with periodic sequence, are derived. In the DCO-rate model, all noise sources are uncorrelated upsampled to the DCO frequency, with the decorrelation sequence allowing the omission of the beat frequency of different noise sources. As a result, the computation resource consumption is significantly reduced. The proposed modeling method is verified against Simulink transient simulations, demonstrating high accuracy in predicting the phase noise.

{\appendix[Intuitive Explanation of $w_{\text{U},L}$]

\begin{figure*}[!t]
\centering
\includegraphics[width=7.14in]{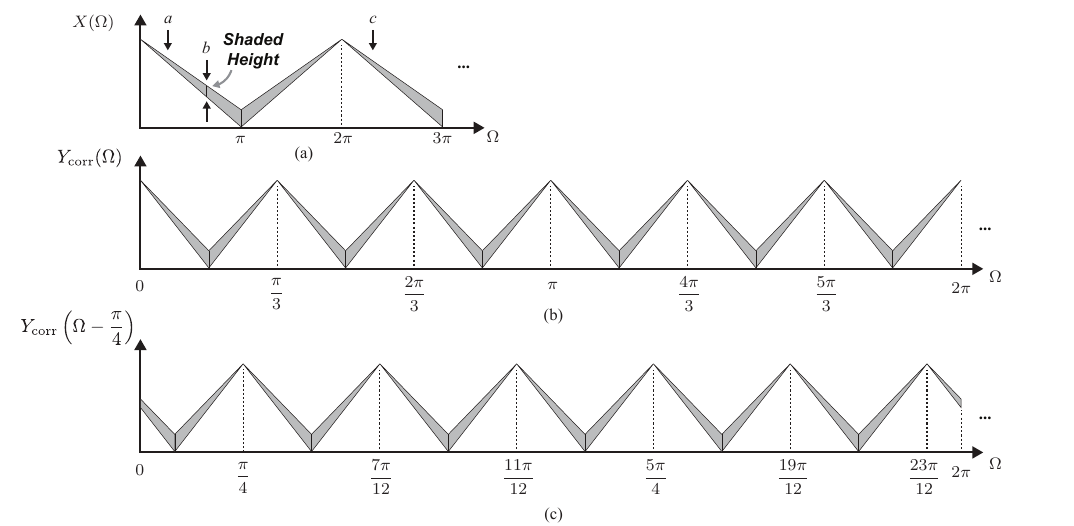}
\caption{Illustration of spectrum with sampling rate conversion. (a) The input spectrum at low sampling rate, (b) the spectrum after upsampling, and (c) the frequency-shifted spectrum after upsampling.}
\label{fig:appen_decorr}
\end{figure*}

This Appendix provides the intuition that helps finding the sequence $w_{\text{U},L}[n]$. In Fig. \ref{fig:block_diagram_math_pre}, $x[m]$ is a WSS low sampling rate sequence that needs to be upsampled. The DTFT of $x[m]$, $X(\Omega)$, is shown in Fig. \ref{fig:appen_decorr}(a). The height inside the shaded area at a specific $\Omega$ contains information on correlation; having different shaded height means uncorrelated spectrum. For example, $X(a)$ is uncorrelated with $X(b)$, i.e., $E\{X(a)X(b)\}=0$, because they have different shaded height. On the other hand, $X(a)$ is correlated with $X(c)$, i.e., $E\{X(a)X(c)\}\neq0$, for having the same shaded height. These shades are only for illustrating the correlation, not the actual value of these spectra. When $x[m]$ is a real WSS random process, $X(\Omega)$ is conjugate symmetric w.r.t. $\Omega = k\pi$ and aperiodic for $\Omega \in [0,2\pi)$.

After upsampling by $L$ times by inserting $L-1$ zeros between adjacent samples, $y_\text{corr}[n]$ is no longer a WSS random process, but a cyclostationary random process. The autocorrelation sequence of $y_\text{corr}[n]$, $R_{yy,\text{corr}}$, is $L$-periodic, making the period of $Y_\text{corr}(\Omega)$ equal to $2\pi/L$. Fig. \ref{fig:appen_decorr} shows $Y_\text{corr}(\Omega)$ when $L=6$.

When $y_\text{corr}[n]$ passes through a system with fundamental frequency $2\pi/N$, the output spectrum of the system $Y_\text{sys}(\Omega)$ is a summation of frequency-shifted $Y_\text{corr}(\Omega)$:
\begin{equation}
Y_\text{sys}(\Omega) = \sum_{k=0}^{N-1} H_{0,k}(\Omega)Y_\text{corr}\left(\Omega-\frac{2\pi k}{N}\right).
\label{eqn:spectrum_out}
\end{equation}

\begin{figure}[!t]
\centering
\includegraphics[width=3.49in]{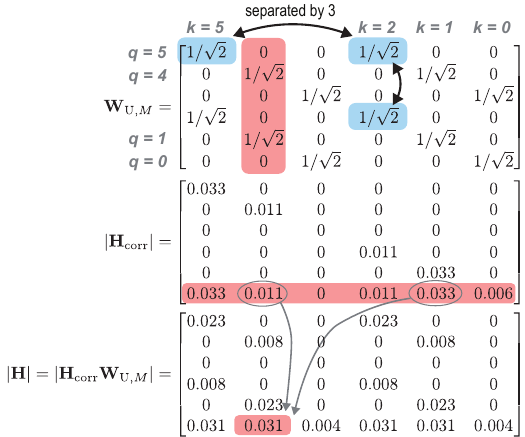}
\caption{Example of the DSM conversion matrix $\mathbf{H}_\text{DSM,tot}$ at 10 kHz for $N=6$ and $M=2$ and its decorrelation conversion matrix $\mathbf{W}_{\text{U},M}$.}
\label{fig:decorr_intuition}
\end{figure}

Fig. \ref{fig:appen_decorr}(c) shows $Y_\text{corr}(\Omega-\pi/4)$ for $N=8$ and $k = 1$. By inspection, $Y_\text{corr}(\Omega-\pi/4)$ is uncorrelated with $Y_\text{corr}(\Omega)$. In general, there are 2 types of frequency shifts:
\begin{enumerate}
    \item \textit{Correlated shift}: the shifted spectrum is correlated with the original spectrum. For $L=6$ and $N=8$, $k_\text{cs}=0,4$. 
    \item \textit{Uncorrelated shift}: the shifted spectrum is uncorrelated with the original spectrum. For $L=6$ and $N=8$, $k_\text{us}=1,2,3,5,6,7$.
\end{enumerate}
Since \eqref{eqn:psd_out} will be used to calculate the output PSD, $Y_\text{corr}(\Omega)$ needs to be pre-decorrelated before feeding into the system. By inspection, the sequence of correlated shift is
\begin{equation}
k_\text{cs}[n] = \begin{cases}
1,& \text{if mod}(n,N/\mathrm{gcd}(L,N))=0 \\ 
{0},&{\text{else}} 
\end{cases}.
\label{eqn:corrshift}
\end{equation}
Therefore the decorrelation sequence in \eqref{eqn:decorrelationwindow} is chosen to be the ``complement'' of the correlated shift sequence \eqref{eqn:corrshift}, and the magnitude is set to preserve the power after decorrelation.

Fig. \ref{fig:decorr_intuition} shows the conversion matrix of the decorrelation sequence $\mathbf{W}_{\text{U},M}$ for $N=6$ and $M=2$. All non-zero elements are separated by 3 circularly in columns and rows, which equals to the period of the correlated shift sequence $k_\text{cs}[n]$. Multiplying $\mathbf{W}_{\text{U},M}$ on the right side of the DSM conversion matrix before decorrelation $\mathbf{H}_\text{corr}$ provides a correction factor on its bottom row, which compensates the underestimated conversion gain caused by ignoring cross-correlated spectrum and using time-averaged PSD of a cyclostationary input. 

\section*{Acknowledgments}
This work is supported by Semiconductor Research Corporation Task No. 3160.040 through UT Dallas' Texas Analog Center of Excellence (TxACE).

}

\bibliographystyle{IEEEtran}
\bibliography{LocalLib}

\end{document}